\documentclass[aps,prl,reprint,groupedaddress,citeautoscript,longbibliography,twocolumn,amsmath]{revtex4-2}

\usepackage{bbold}
\usepackage{braket}
\usepackage{tensor}
\usepackage{parskip}

\usepackage{graphicx}

\usepackage{xcolor}
\usepackage{color,soul}

\sethlcolor{white}
\newcommand{\highlight}[1]{\colorbox{white}{$\displaystyle#1$}}

\begin{document}

\title{Light-Hole Gate-Defined Spin-Orbit Qubit}

\author{Patrick Del Vecchio}
\author{Oussama Moutanabbir}
\email{oussama.moutanabbir@polymtl.ca}
\affiliation{Department of Engineering Physics, \'Ecole Polytechnique de Montr\'eal, Montr\'eal, C.P. 6079, Succ. Centre-Ville, Montr\'eal, Qu\'ebec, Canada H3C 3A7}

\date{\today}

\begin{abstract}
The selective confinement of light-holes (LHs) is demonstrated by introducing a low-dimensional system consisting of highly tensile-strained Ge quantum well enabling the design of an ultrafast gate-defined spin qubit under the electric dipole spin resonance. The qubit size-dependent $g$-factor and dipole moment are mapped, and the parameters inducing their modulation are discussed. It is found that the LH qubit dipole moment is 2 to 3 orders of magnitude higher than that of the canonical heavy-hole qubit. This behavior originates from the significant spin splitting resulting from the combined action of large cubic and linear Rashba spin-orbit interactions that are peculiar to LHs. The qubit relaxation rate is also affected by the strong spin-orbit interaction and follows typically a $B^7$ behavior. The proposed all-group IV, direct bandgap LH qubit provides an effective platform for a scalable qubit-optical photon interface sought-after for long-range entanglement distribution and quantum networks.

\end{abstract}

\maketitle

\footnotetext[1]{See Supplemental Material at link.aps.org for details on the $k\cdot p$ Hamiltonian, the subband edges basis, the subband parameters in the LH effective Hamiltonian, the solution of the QD Hamiltonian, the LH qubit relaxation rate and the GeSn alloy material parametrization.}





Gated quantum dots (QDs) exploiting the strong spin-orbit interaction (SOI) of holes and their quiet quantum environment provide practical building blocks for quantum processors ~\cite{Hendrickx2020_1,Hendrickx2020_2,Hendrickx2021,Lawrie2020,Bogan2019,Wang2016,Watzinger2018,Vukusic2018,Jirovec2022,Wang2022,Bosco2021}. However, due to the restricted choice of low-dimensional systems (e.g., Ge/SiGe), current hole spin qubits are based predominately on heavy-hole (HH) spins ~\cite{Scappucci2020}. Notwithstanding this progress, the ability to utilize light-hole (LH) spins would enable additional degrees of freedom to engineer qubits with extended functionalities. Indeed, LHs allow simple schemes for a direct mapping of superposition from a flying qubit to a stationary spin qubit~\cite{Vrijen2001} as well as a better resilience against charge noise ~\cite{AbadilloUriel2017} and an enhanced proximity-induced superconductivity transfer~\cite{Moghaddam2014}. Additionally, LHs are also known to have strong SOI yielding fast Rabi oscillations~\cite{AbadilloUriel2017}. Nevertheless, the development of LH qubits has been hampered by the lack of proper material systems. Here, we address this limitation and introduce a new low-dimensional system to control LH states.

The selective confinement of LHs in Ge quantum well (QW) requires sufficiently high tensile strain, which can be achieved using the emerging germanium-tin (Ge$_{1-x}$Sn$_{x}$) alloys~\cite{Assali2022}. Ge/Ge$_{1-x}$Sn$_{x}$ hole spin devices combine all advantages that are inherent to group IV semiconductors~\cite{Hollmann2020,Zwerver2022}. Besides the weaker hyperfine interaction with the surrounding nuclear spin bath resulting from the p-symmetry of the hole wavefunction ~\cite{Testelin2009,Machnikowski2019,Philippopoulos2020}, the strong SOI in the valence band of Ge and Sn would enable all-electrical driving of the qubit without the need for an external RF transmission line and create rich spin-related phenomena unique to holes~\cite{Moriya2014,DelVecchio2020}. Moreover, the Ge$_{1-x}$Sn$_{x}$ alloy spans a wide range of lattice parameters~\cite{Moutanabbir2021,Polak2017,Madelung1991}, which is useful to control the hole spin properties through the epitaxial strain directly on silicon wafers~\cite{Stange2016,Atalla2022,Assali2022}.

\begin{figure*}[th]
    \centering
    \includegraphics[scale=0.145]{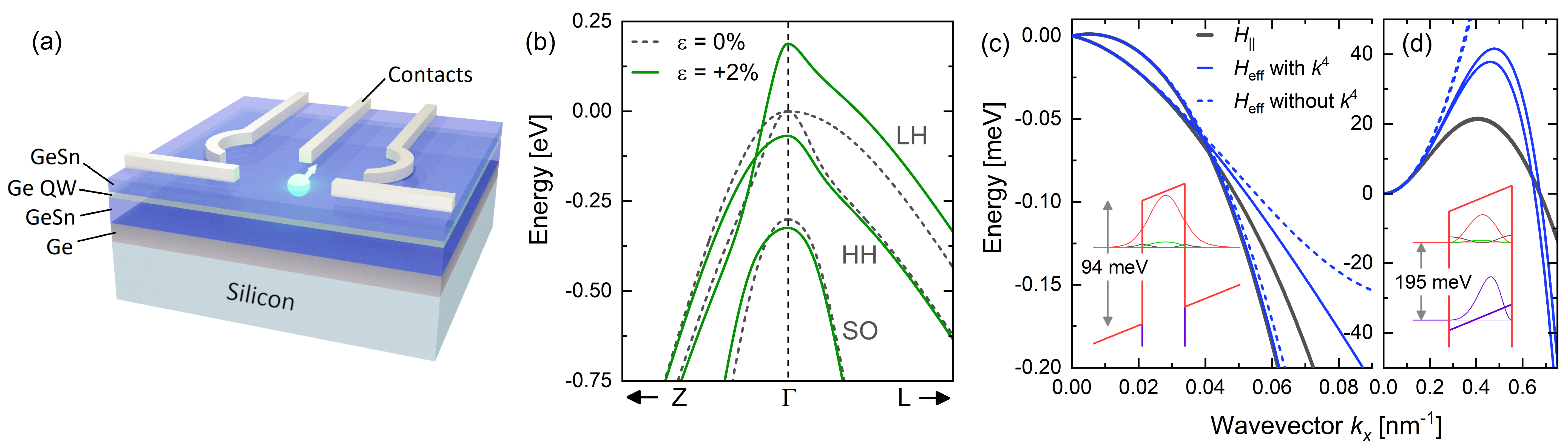}
    \caption{(a) Schematic of a gate-defined Ge/Ge$_{1-x}$Sn$_{x}$ LH qubit. (b) Band structure of the valence band in bulk Ge without strain (dashed lines) and with $2\%$ tensile strain (solid lines). $k\cdot p$ parameters taken from Ref~\cite{Rideau2006}. (c)-(d) Ground LH subband dispersion from the numerical diagonalization of $H_\parallel$ (black solid lines) and from $H_\text{eff}$ with (without) $k^4$ terms in solid (dashed) blue. (c)~: Ge/Ge$_{1-x}$Sn$_{x}$ LH QW with $E_z = 1\,\text{MV}/\text{m}$. (d)~: Infinite Ge LH QW with $E_z = 5\,\text{MV}/\text{m}$. Energy scale is in meV in both panels. Insets show the envelope probability density of the lowest subband (ground HH subband is also shown in (d)). The larger component of the wavefunction is the LH part of the spinor (red). The smaller component with one lobe corresponds to the SO part (green) and the component with two small lobes at the Ge interfaces corresponds to the CB part (black). The tensile strain in Ge is $2.38\%$ in both cases corresponding to $\textit{x} = 0.15$.}
    \label{fig:disp}
\end{figure*}

Fig. \ref{fig:disp} illustrates the gate-defined LH QD. Note that the lattice mismatch between Ge and Ge$_{1-x}$Sn$_{x}$ induces a significant tensile strain in the Ge layer, which lifts the HH-LH degeneracy yielding a LH-like valence band edge (Fig \ref{fig:disp}b). The Ge$_{1-x}$Sn$_{x}$/Ge/Ge$_{1-x}$Sn$_{x}$ heterostructure confines LHs in the Ge layer for $x$ typically higher than $0.11$, while the HHs are pulled into the Ge$_{1-x}$Sn$_{x}$ barriers~\cite{Assali2022}. A set of electrostatic gates on top of the heterostructure helps confine the LH in the plane by applying a DC voltage. Note that Ge becomes a direct bandgap semiconductor at a tensile strain higher than  $1.8\%$. The EDSR is performed by applying a microwave voltage. A feature that is sometimes neglected~\cite{Terrazos2021,Bulaev2007,Wang2021} but needs to be accounted for in this system is the spread of the LH wavefunction into the barriers. Because the HHs are located in the barriers, LH-HH mixing wavefunction overlap only occurs outside the QW. Assuming a hard wall potential at the interface is therefore equivalent to neglecting entirely the LH-HH mixing. Moreover, the LH subband dispersion non-parabolicity must also be considered. The theoretical framework below for the in-plane motion of the LHs explicitly takes into account these peculiar features. 

Eight-band $k\cdot p$ theory~\cite{Kane1957} is used for the derivation of an effective Hamiltonian for the 2D LH gas incorporating the Bir-Pikus Hamiltonian and thus the effects of bi-axial epitaxial strain~\cite{Bir1974}. $[001]$-oriented substrates are considered, with the growth direction parallel to the $z$-axis. The operator ordering between material parameters and wavevector components require special care to avoid spurious solutions and to properly include the effects of an external magnetic field~\cite{Foreman1997,Eissfeller2012}. An out-of-plane magnetic field $\mathbf{B} = B\mathbf{e}_z$ results in the following commutation relations for the mechanical wavevector components~: $[K_\alpha,K_\beta] = \epsilon_{\alpha\beta,z}/i\lambda^2$, where $\epsilon$ is the Levi-Civita tensor, $\lambda = \sqrt{\hbar/eB}$ is the magnetic length and $\alpha,\beta = \{x,y,z\}$. The mechanical wavevector $\mathbf{K} = \mathbf{k} + e\mathbf{A}/\hbar$ is given in terms of the canonical wavevector $\mathbf{k}\to -i\nabla$. In the symmetric gauge, the vector potential $\mathbf{A} = B/2(-y\mathbf{e}_x + x\mathbf{e}_y)$.

The total Hamiltonian $H_\parallel$ for the in-plane motion of holes and electrons is written as a sum of different contributions \cite{Eissfeller2012}~: $H_\parallel = H_{k\cdot p}(\mathbf{K}_\parallel;k_z) + V(z)$, where $\mathbf{K}_\parallel = K_x\mathbf{e}_x + K_y\mathbf{e}_y$, $H_{k\cdot p}$ is the eight-band $k\cdot p$ matrix including strain and magnetic effects~\cite{Kane1957,Eissfeller2012,Note1} and $V(z)$ is the band alignment. This last term also includes the effects of an out-of-plane electric field $\mathbf{E} = E_z\mathbf{e}_z$. The first step to find an effective LH Hamiltonian is to calculate the envelope functions and energies of $H_\parallel$ at $K_x = K_y = 0$ and $B = 0$. This provides an orthonormal basis (a set of subband edges) on which $H_\parallel$ is projected at finite $\mathbf{K}_\parallel$ and $B > 0$. This orthonormal basis contains two types of subbands. The first are pure HH subbands ($\text{H}$ subbands) and the second are superpositions of LH, SO holes, and CB electrons ($\eta$ subbands)~:

\begin{subequations}
\begin{gather}
  \Ket{\text{H};l,\sigma} = \Ket{\frac{3}{2},\frac{3\sigma}{2}}\Ket{l}_h \\
  \Ket{\eta;j,\sigma} = \Ket{\frac{1}{2},\frac{\sigma}{2}}_c\Ket{j}_c + \Ket{\frac{3}{2},\frac{\sigma}{2}}\Ket{j}_\ell + \sigma\Ket{\frac{1}{2},\frac{\sigma}{2}}\Ket{j}_s,
\end{gather}
\end{subequations}

where $l$ and $j$ are respectively the subband indices for $\text{H}$ and $\eta$ subbands and $\sigma = \pm 1$ is the pseudo-spin index. The first ket in each term represents bulk Bloch functions at the $\Gamma$ point, while the second ket represents the envelope functions. The labels $h,c,\ell,s$ refer to the HH, CB, LH and SO part of the spinor, respectively. Because subbands are either of type $\text{H}$ or $\eta$, a ``LH'' subband is understood as an $\eta$ subband such that $\tensor[_\ell]{\Braket{j|j}}{_\ell} > \tensor[_c]{\Braket{j|j}}{_c}$ and $\tensor[_\ell]{\Braket{j|j}}{_\ell} > \tensor[_s]{\Braket{j|j}}{_s}$. Following the projection of $H_\parallel$ upon the basis $\{\Ket{\eta},\Ket{\text{H}}\}$, a 4th order Schrieffer-Wolff transformation~\cite{Winkler2003} is applied leading to an effective Hamiltonian for $\eta$ subbands~:

\begin{equation}\label{eff}
  \begin{split}
    H_{\text{eff}} &= \alpha_0\tilde{\gamma}K_\parallel^2 + \frac{\alpha_0}{\lambda^2}\frac{\tilde{g}}{2}\sigma_z + \alpha_0^2\tilde{\gamma}'K_\parallel^4 + \frac{\alpha_0^2}{\lambda^4}\tilde{g}' +\frac{\alpha_0^2}{\lambda^2}\gamma_\lambda K_\parallel^2\sigma_z \\
    &+ \alpha_0^2\left[\left(\zeta K_-^4 + h.c.\right)\sigma_+\sigma_- + \left(\zeta K_+^4 + h.c.\right)\sigma_-\sigma_+\right] \\
    &+ i\beta_1\left(K_-\sigma_+-K_+\sigma_-\right) - i\beta_2\left(K_+^3\sigma_+-K_-^3\sigma_-\right) \\
    &+ i\beta_3\left(K_-K_+K_-\sigma_+-K_+K_-K_+\sigma_-\right),
  \end{split}
\end{equation}

\begin{figure}[th]
    \centering
    \includegraphics[scale=0.38]{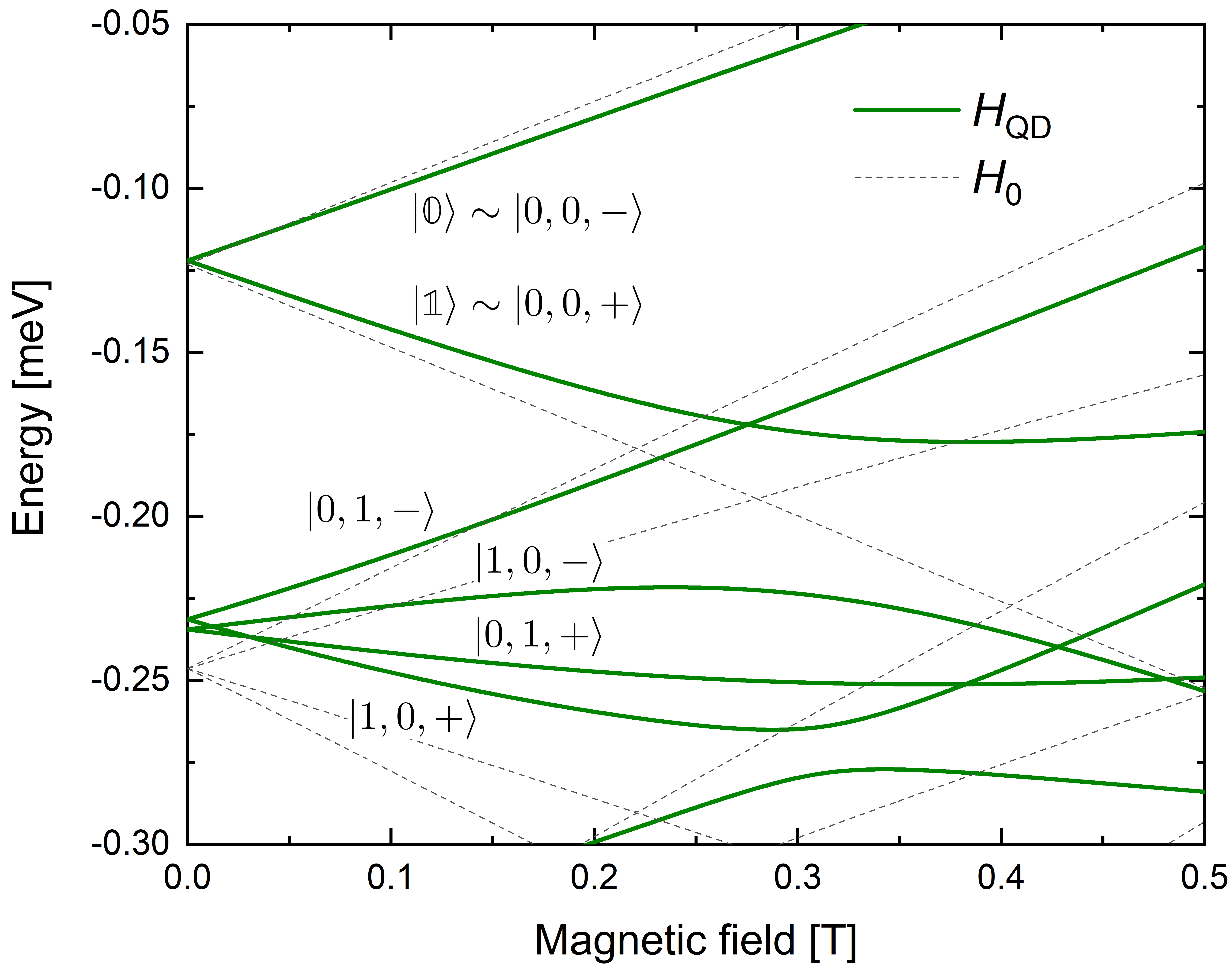}
    \caption{LH QD orbital energies as a function of $B$. The QD radius $r_0$ at $B=0$ is kept constant at $25\,\text{nm}$. Black dotted lines represent the eigenvalues of $H_0$, while the solid green lines are those of $H_\text{QD}$. The qubit levels $\Ket{\mathbb{0}}$ and $\Ket{\mathbb{1}}$ and excited orbitals are displayed with their main contributions from the eigenstates $\Ket{n_1,n_2,\sigma}$ for $B = 0.1\,\text{T}$. The QW parameters are the same as in Fig. \ref{fig:disp}c.}
    \label{fig:orbitals}
\end{figure}

\noindent with $\alpha_0 = \hbar^2/(2m_0)$, $m_0$ the free electron mass, $K_\pm = K_x\pm iK_y$, $\sigma_\pm = (\sigma_x\pm i\sigma_y)/2$ and $\sigma_{x,y,z}$ the Pauli matrices. The first term in \eqref{eff} corresponds to the parabolic contribution in the dispersion relation, with $\tilde{\gamma}$ the effective mass parameter. The second term corresponds to the linear Zeeman splitting, with $\tilde{g}$ the effective $g$-factor. The next three terms correspond to non-parabolicity ($\tilde{\gamma}'$), Zeeman splitting non-linearity ($\tilde{g}'$), and a hybrid term proportional to $K_\parallel^2/\lambda^2$. The band structure anisotropy is taken into account by the $\zeta$ parameter. Finally, the last three terms correspond to the linear Rashba splitting ($\beta_1$) and two kinds of cubic Rashba splitting ($\beta_2$ and $\beta_3$). The effective parameters in \eqref{eff} in terms of the envelopes $\Ket{l}_h$ and $\Ket{j}_{c,\ell,s}$ are presented in~\cite{Note1}. \hl{This approach is similar to that employed by~\mbox{\cite{Bulaev2007,Wang2021,Terrazos2021}} for instance}. However, here the spread of the wavefunction into the barriers and the effects of $\mathbf{E}$ are implicitly taken into account from the shape of the envelopes, \hl{and the effective parameters in {\eqref{eff}} are calculated from a larger subband edge basis due to the large amount of HH levels in the barriers}. Fig. \ref{fig:disp} shows the dispersion of the ground LH subband for two different QWs. Fig.\ref{fig:disp}(c) displays the case of a $13\,\text{nm}$ Ge QW with relaxed Ge$_{0.85}$Sn$_{0.15}$ barriers. Close to $k_\parallel = 0$, $H_\text{eff}$ fits exactly $H_\parallel$ because the effective parameters in \eqref{eff} are given exactly by 4th order perturbation theory. $H_\text{eff}$ diverges from $H_\parallel$ further away from $k_\parallel = 0$ because $k^5$-terms or higher become important. This behavior is exacerbated for a QW with infinite band offsets (Fig.\ref{fig:disp}(d)). In both cases, $H_\text{eff}$ diverges faster when $k^4$ terms are neglected.

The calculations also predict an effective mass for the ground LH subband of the opposite sign (clearly visible in Fig \ref{fig:disp}d) for certain QW parameters. Such behavior was also observed for different material systems ~\cite{Ahn1988,Sanders1987,Hayden1991}. This effect becomes more prominent as the QW thickness decreases. The typical hole-like dispersion is recovered above a critical thickness, corresponding to $11\,\text{nm}$ for the Ge/GeSn QW system in Fig. \ref{fig:disp}c. To simplify the  LH qubit calculations, the thickness is fixed at $13\,\text{nm}$.

The QD Hamiltonian includes the isotropic and parabolic confinement from the top gates~: $H_\text{QD} = H_\text{eff} + m^*\omega_0^2\left(x^2+y^2\right)/2$, where $m^* = m_0/\tilde{\gamma}$ is the in-plane effective mass. $H_\text{QD}$ is diagonalized by first writing $H_\text{QD}= H_0 + H'$, where $H_0$ consists of the first two terms in \eqref{eff} plus the parabolic confinement~:

\begin{align}
  H_0 &= \alpha_0\tilde{\gamma}K_\parallel^2 + \frac{1}{2}m^*\omega_0^2\left(x^2+y^2\right) + \frac{\alpha_0}{\lambda^2}\frac{\tilde{g}}{2}\sigma_z. \\
  \begin{split}
      &\highlight{= \hbar\omega_l\left(a_1^\dag a_1 + a_2^\dag a_2 + 1\right)} \\
      &\highlight{+ \frac{\hbar\omega_c}{2}\left(a_1^\dag a_1 - a_2^\dag a_2\right) + \frac{\alpha_0}{\lambda^2}\frac{\tilde{g}}{2}\sigma_z,}
  \end{split}
\end{align}

\noindent \hl{where}

\begin{align}
    \highlight{a_1} &\highlight{= \frac{x - iy}{2r} + \frac{irk_-}{2},} & \highlight{a_2} &\highlight{= \frac{x + iy}{2r} + \frac{irk_+}{2}}
\end{align}

\noindent \hl{are ladder operators, $k_\pm = k_x\pm ik_y$, $\omega_c = eB/m^*$, $\omega_l^2 = \omega_0^2 + \omega_c^2/4$ and $r = \sqrt{\hbar/(m^*\omega_l)}$ is the effective quantum dot radius~\mbox{\cite{Note1}}}. The eigenstates of $H_0$, the so-called Fock-Darwin orbitals $\Ket{n_1,n_2,\sigma}$ with $n_{1,2} = 0,1,\dots$ and $\sigma=\pm 1$, provide an orthonormal basis on which $H_\text{QD}$ is projected. The eigenvalues of the resulting matrix for $H_\text{QD}$ in the Fock-Darwin basis are then solved numerically. The two lowest energy orbitals $\ket{\mathbb{0}}$ and $\ket{\mathbb{1}}$ corresponding to energies $E_\mathbb{0}$ and $E_\mathbb{1}$ define the qubit. These are mostly composed of the Fock-Darwin orbitals $\ket{0,0,-}$ and $\ket{0,0,+}$ respectively, plus higher-energy orbitals~:

\begin{subequations}
\begin{align}
    \ket{\mathbb{0}} &= \ket{0,0}\ket{-} + \left(c_{0,1}^{(\mathbb{0})}\ket{0,1} + c_{3,0}^{(\mathbb{0})}\ket{3,0}+c_{1,2}^{(\mathbb{0})}\ket{1,2}\right)\ket{+} \\
    \ket{\mathbb{1}} &= \ket{0,0}\ket{+} + \left(c_{1,0}^{(\mathbb{1})}\ket{1,0} + c_{0,3}^{(\mathbb{1})}\ket{0,3}+c_{2,1}^{(\mathbb{1})}\ket{2,1}\right)\ket{-}.
\end{align}
\end{subequations}

\noindent The coefficients $c_{n_1,n_2}^{(\mathbb{0},\mathbb{1})}$ were extracted from the numerical diagonalization of $H_\text{QD}$ to avoid artifacts near crossings between Fock-Darwin orbitals. They can be evaluated with perturbation theory away from these crossings~\cite{Note1}. For a driving field $\tilde{\mathbf{E}}(t) = \mathbf{e}_xE_\text{AC}\cos(\omega t)$, where $\hbar\omega = |E_\mathbb{0} - E_\mathbb{1}|$ is the qubit energy, the Rabi frequency $\Omega$ is given in terms of the qubit dipole moment $d=e\Braket{\mathbb{0}|x|\mathbb{1}}$ by $\Omega = E_\text{AC}|d|/\hbar$.

Fig. \ref{fig:orbitals} shows the QD orbital energies as a function of the out-of-plane magnetic field for a QD radius $r_0 = \sqrt{\hbar/m^*\omega_0} = 25\,\text{nm}$ and the same QW parameters as in Fig. \ref{fig:disp}a. The qubit undergoes a transition from a spin qubit to a charge qubit at the crossing between $\Ket{\mathbb{1}}$ and the mostly $\Ket{0,1,-}$ orbital near $B = 0.275\,\text{T}$. The two levels cross because $\Ket{0,1,-}$ is not present in the expansion of $\Ket{\mathbb{1}}$.

\begin{figure}[th]
    \centering
    \includegraphics[scale=0.36]{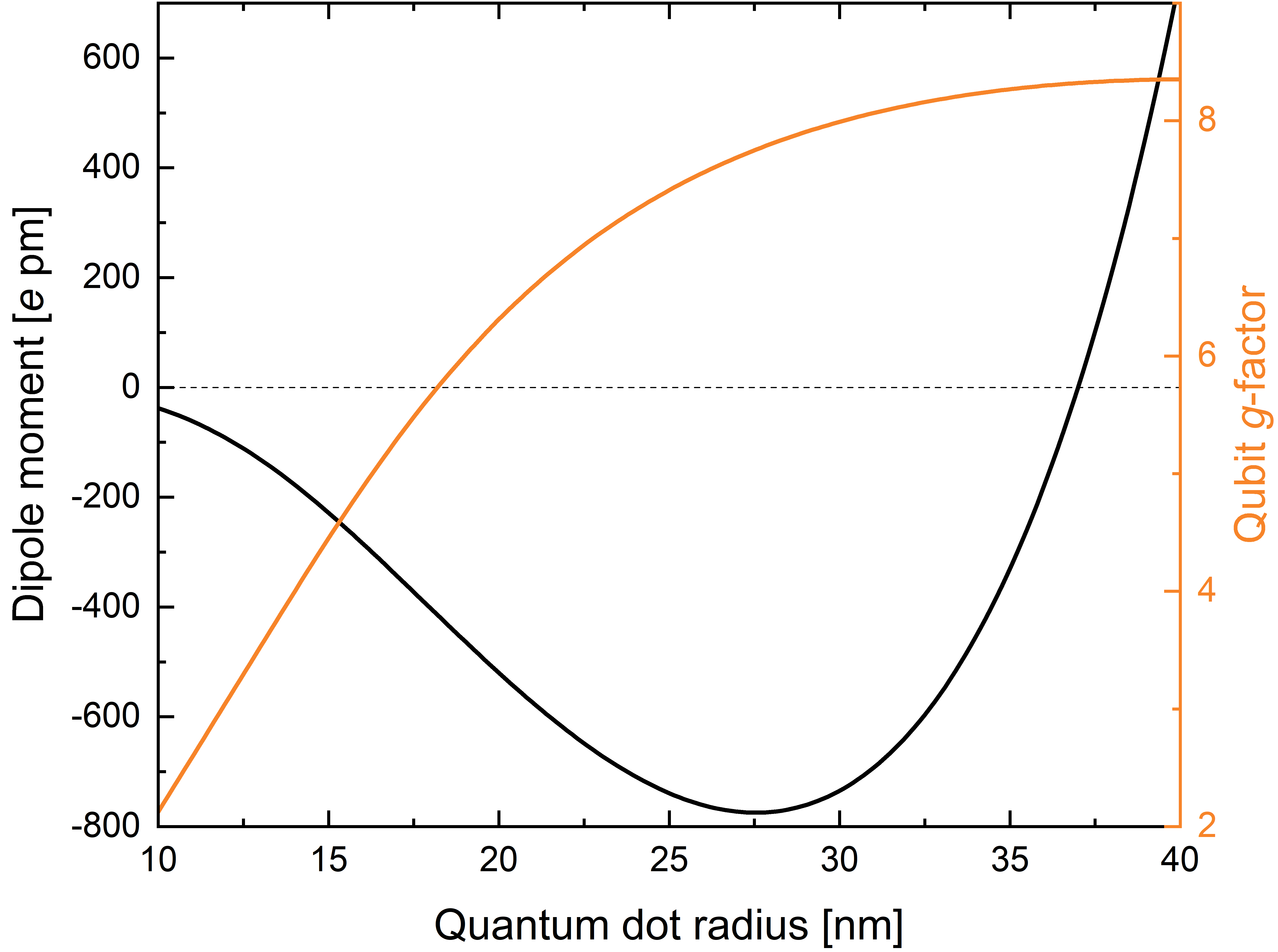}
    \caption{LH qubit dipole moment (left) and absolute value of the $g$-factor (right) as a function of the QD radius $r_0$. The magnetic field is fixed at $0.05\,\text{T}$. The QW parameters are the same as in Fig. \ref{fig:disp}c.}
    \label{fig:qubit}
\end{figure}

The qubit dipole moment $d$ and the qubit $g$-factor $|g_\text{QD}|$ are plotted in Fig. \ref{fig:qubit} as a function of the QD radius $r_0$ and $B = 0.05\,\text{T}$. The $g$-factor has a strong dependence on $r_0$ for small radii, and approaches asymptotically the QW value $\tilde{g} = 8.69$ at large $r_0$. The large $\tilde{g}$ value originates from the first order approximation $\tilde{g} \approx 2\kappa$ for a LH spin in a perpendicular magnetic field with $\kappa = 3.41$ in Ge. Deviations from $2\kappa$ come from the spread of the wavefunction and 2nd order corrections~\cite{Note1,Lyanda-Geller2022}. The dipole moment $d$ takes very large values for two main reasons. First, the coefficient $\beta_3$ that contributes to EDSR by introducing a $\Ket{1,0,-}$ contribution into $\Ket{\mathbb{1}}$ depends on the sum $(\gamma_2 + \gamma_3)$ for LHs~\cite{Note1} whereas for HHs it depends on the difference $(\gamma_2 - \gamma_3)$. In Ge, $\gamma_2 \approx \gamma_3$~\cite{Paul2016} and therefore $\beta_3$ is much larger for LHs. Secondly, LHs are subject to a linear Rashba spin splitting proportional to $\beta_1$, which is non-existent for HHs. This additional term contributes to a large $d$ similarly to $\beta_3$ by increasing the contribution of $\Ket{1,0,-}$ into $\Ket{\mathbb{1}}$. At $\approx27.5\,\text{nm}$, $|d|$ reaches a maximum as a result of the combined effects of $\beta_1$ and $\beta_3$, which gives a dipole moment that is $2$ to $3$ orders of magnitude larger than that of HHs in a compressively strained Ge~\cite{Terrazos2021,Venitucci2019}. \hl{For instance, an in-plane driving field as small as $E_\text{AC} = 1\,\text{mV}/\mu\text{m}$ gives a Rabi frequency $\Omega\approx 1.2\,\text{GHz}$.}

There is, however, a range of QD radii where the dipole moment is very small (Fig. \ref{fig:qubit}). This happens because both $\beta_1$ and $\beta_3$ contribute to $d$. When $B$ is small such that $\Ket{\mathbb{1}}$ is far from the excited orbitals, the dipole moment is given by

\begin{equation}\label{dipole}
    |d| \approx \frac{em^{*2}r^2|\tilde{g}|\mu_\text{B}B}{\hbar^4}\left|r^2\beta_1+2\beta_3\right|.
\end{equation}

\noindent Therefore, when $\beta_1\beta_3 < 0$, $d$ can vanish at specific values of $r_0$ and $B$. For the QW parameters in Fig. \ref{fig:disp}a, $\beta_1 = 0.42\,\text{meV}\,\text{nm}$ and $\beta_3 = -290\,\text{meV}\,\text{nm}^3$ which causes the dipole moment to vanish at $B = 0.05\,\text{T}$ and $r_0\approx 37\,\text{nm}$. For $r_0 > 37\,\text{nm}$ $d$ increases again, but at the cost of a smaller orbital energy spacing.

An important feature of LH qubits is that EDSR is driven by both $\eta$-H and $\eta$-$\eta$ mixing. This is because there is an allowed 1st order coupling between $\eta$ subbands $\Braket{\eta;j,+|H_\parallel|\eta;j',-} = R_{j,j'}K_-$ that is nonexistent for HHs~\cite{Note1}. The $\eta$-H mixing part contributes mainly to the $\beta_3$ parameter through a term proportional to $(\gamma_2 + \gamma_3)$, while $\eta$-$\eta$ mixing contributes to both $\beta_3$ and $\beta_1$. Notably, these two types of mixing are of equal importance given that $\beta_1$ and $\beta_3$ can interfere to suppress the dipole moment (c.f. \eqref{dipole}).

\begin{figure}[t]
    \centering
    \includegraphics[scale=0.38]{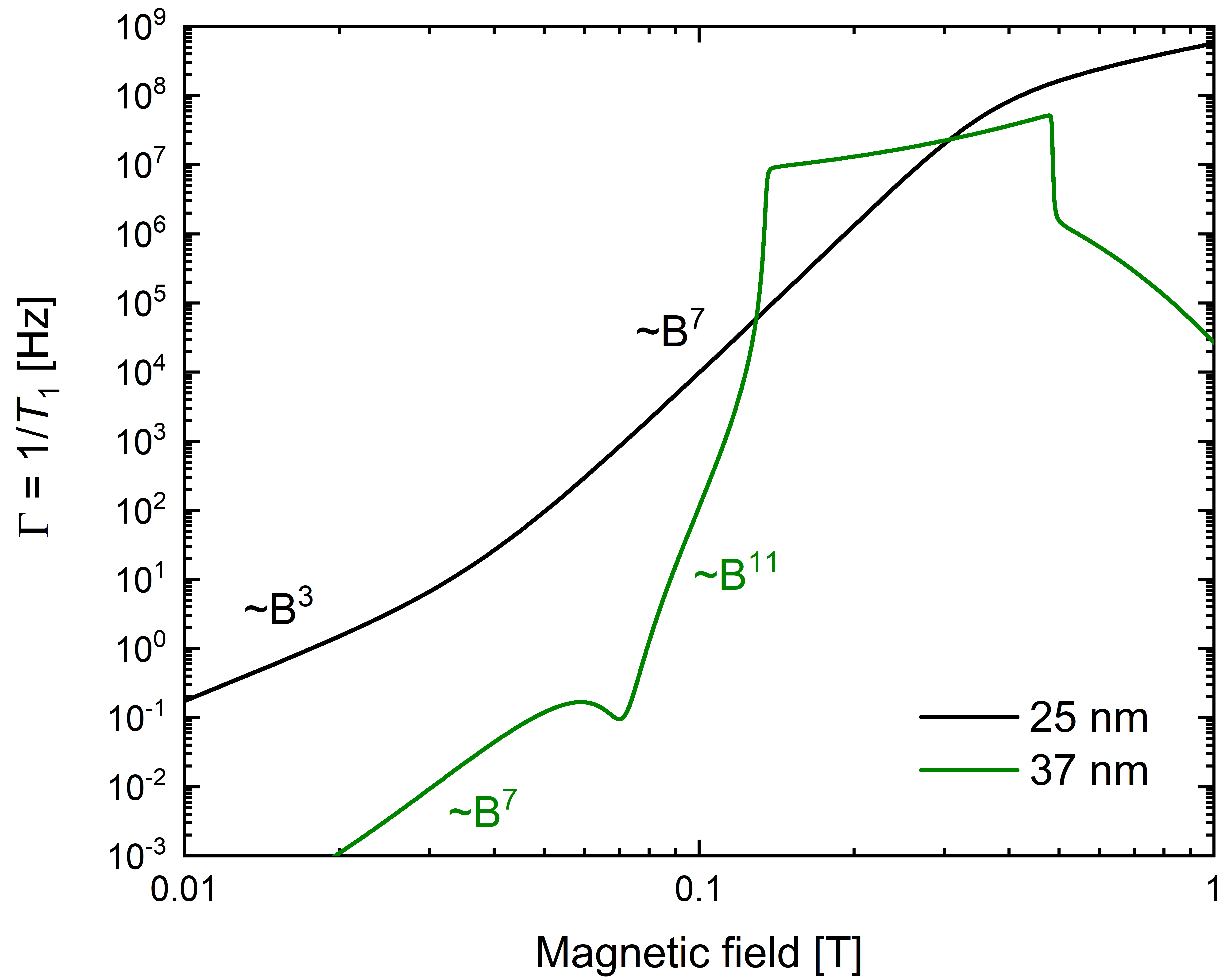}
    \caption{Relaxation rate $\Gamma = 1/T_1$ as a function of the magnetic field. The QW parameters are the same as in Fig. \ref{fig:disp}c.}
    \label{fig:rate}
\end{figure}

The relaxation time $T_1 = 1/\Gamma$ of the LH qubit was also evaluated for the system in Fig. \ref{fig:disp}a. The coupling of the hole to acoustic phonons was considered. The total relaxation rate $\Gamma = \Gamma_\text{em} + \Gamma_\text{abs}$, where $\Gamma_\text{em}$ ($\Gamma_\text{abs}$) is the rate associated with the emission (absorption) of one phonon. Each of these rates is calculated by Fermi's golden rule~:

\begin{equation}
    \Gamma_i = \frac{2\pi\mathcal{V}}{\hbar}\sum_\alpha{\int{\frac{d^3q}{8\pi^3}\left|\Braket{f|W_\alpha|i}\right|^2\delta(\hbar\omega - \hbar\omega_{\alpha\mathbf{q}})}},
\end{equation}

\noindent where $\mathcal{V}$ is the volume of the system and $\hbar\omega_{\alpha\mathbf{q}} = \hbar v_\alpha q$ is the phonon energy in branch $\alpha=\{\text{LA,TA1,TA2}\}$ and with momentum $\mathbf{q}=q\hat{\mathbf{q}}$. $\Ket{i,f}$ represent initial and final qubit levels upon absorption or emission of a phonon. The operator $W_\alpha$ is derived from the hole-phonon Hamiltonian in a procedure similar to that in~\cite{Li2020,Wang2021,Bulaev2005PRB,Woods2004}. See~\cite{Note1} for details. Importantly, the matrix element $\Braket{f|W_\alpha|i}$ takes into account the relaxation rate associated with all three spin-orbit parameters $\beta_1$, $\beta_2$, and $\beta_3$.

Fig. \ref{fig:rate} shows the computed $\Gamma$ as a function of $B$ for the Ge/GeSn QW system in Fig. \ref{fig:disp}a at $r_0 = 25\,\text{nm}$. A relaxation time $T_1 = 100\,\mu\text{s}$ was extracted at $B = 0.1\,\text{T}$. Moreover, $\Gamma$ follows a $B^7$ behavior when $B\gg \sqrt{12}k_\text{B}T/(g_\text{QD}\mu_\text{B})$ and a $B^6$ behavior when $B\ll \sqrt{12}k_\text{B}T/(g_\text{QD}\mu_\text{B})$. This higher relaxation rate for LHs compared to HHs~\cite{Wang2021,Scappucci2020} is due to the larger spin-orbit coupling parameters $\beta_{1,2,3}$. The $B^7$ behavior at low temperature is associated to the spin-orbit term $\left|r^2\beta_1+2\beta_3\right|$ that was encountered in eq. \eqref{dipole} and from the sum $c_{0,1}^{(\mathbb{0})}+c_{1,0}^{(\mathbb{1})}\sim B$.

Similar calculations were also performed at a QD radius $r_0=37.9\,\text{nm}$ for which the dipole moment vanishes at $B = 0.05\,\text{T}$ (Fig. \ref{fig:rate}). In this case, two different regimes were observed~: for $B \ll 0.05\,\text{T}$ the relaxation rate exhibits a $B^7$ behavior, but at $B\gg 0.05\,\text{T}$ it evolves as $\sim B^{11}$. This is because the term associated with $\left|r^2\beta_1+2\beta_3\right|$ vanishes and the dominating terms in $\Gamma$ are those associated with $\beta_2$, $\beta_3$ alone and the superposition coefficients $c_{n_1,n_2}^{(\mathbb{0},\mathbb{1})}$ with $n_1+n_2=3$. At $B = 0.1\,\text{T}$, $T_1 = 8\,\text{ms}$, which is consistent with a much smaller dipole moment at this radius. The abrupt change in behavior around $B = 0.14\,\text{T}$ is due to a very small anti-crossing between $\ket{\mathbb{1}}$ and the mostly $\ket{1,0,-}$ orbital, while at $B = 0.5\,\text{T}$ it is caused by a small anti-crossing between the mostly $\ket{1,0,-}$ and the mostly $\ket{0,3,-}$ orbital.

In conclusion, this work unravels the spin properties of a light-hole gated quantum dot in tensile strained Ge under EDSR. A detailed framework is described taking into account the spread of the envelopes in the barriers surrounding the quantum well and the effects of the dispersion non-parabolicity. It was found that light-holes have a dipole moment $d$ significantly larger than that of the heavy-holes due to a larger cubic Rashba parameter ($\beta_3$) and the existence of a non-zero linear Rashba parameter ($\beta_1$). Interestingly, $\beta_1$ and $\beta_3$ can interfere destructively and cause the dipole moment to vanish at a specific quantum dot size. The relaxation rate $\Gamma$ of a light hole qubit follows a $B^7$ behavior, except when $d\approx 0$ where $\Gamma$ follows a $B^{11}$ behavior. This direct bandgap Ge/GeSn device structure provides additional degrees of freedom to implement silicon-compatible and scalable quantum processors leveraging the advantages of light-hole spin properties in addition to their efficient coupling with optical photons and their ability to transfer superconductivity.  

\subsection*{Acknowledgment}

O.~M. acknowledges support from NSERC Canada, Canada Research Chairs, Canada Foundation for Innovation, Mitacs, PRIMA Qu\'ebec, Defence Canada (Innovation for Defence Excellence and Security, IDEaS), and NRC Canada (New Beginnings Initiative). 

%

\end{document}